\documentclass[conference]{IEEEtran}
\IEEEoverridecommandlockouts
\usepackage{cite}
\usepackage{amsmath,amssymb,amsfonts}
\usepackage{algorithmic}
\usepackage{graphicx}
\usepackage{textcomp}
\usepackage{xcolor}
\usepackage{diagbox}
\usepackage{multirow}
\def\BibTeX{{\rm B\kern-.05em{\sc i\kern-.025em b}\kern-.08em
    T\kern-.1667em\lower.7ex\hbox{E}\kern-.125emX}}

\usepackage{fancyhdr}
\fancypagestyle{firstpage}{
    \fancyhf{} % Clear all header and footer fields
    \fancyhead[L]{\small 2024 27th International Conference on Computer and Information Technology (ICCIT)\\
    20-22 December 2024, Cox’s Bazar, Bangladesh}
    \fancyfoot[L]{\small 979-8-3315-1909-4/24/\$31.00~\copyright2024 IEEE}
}

\begin{document}

\title{Enhancing Transfer Learning for Medical Image Classification with SMOTE: A Comparative Study\\

}

% \author{
% \IEEEauthorblockN{Md. Zehan Alam}
% \IEEEauthorblockA{\textit{Department of Mathematics} \\
% \textit{University of Dhaka}\\
% Dhaka, Bangladesh \\
% zehan.alamm@gmail.com}
% \and
% \IEEEauthorblockN{Tonmoy Roy}
% \IEEEauthorblockA{\textit{Department of Data Analytics} \\
% \textit{and Information Systems} \\
% \textit{Utah State University}\\
% Logan, Utah, USA \\
% tonmoy.roy@usu.edu}
% \and
% \IEEEauthorblockN{H.M. Nahid Kawsar}
% \IEEEauthorblockA{\textit{Department of Applied} \\
% \textit{Mathematics} \\
% \textit{Noakhali Science and}\\
% \textit{Technology University}\\
% Noakhali, Bangladesh \\
% nahidkawsarnhd3@gmail.com}
% \and
% \IEEEauthorblockN{Iffat Firozy Rimi}
% \IEEEauthorblockA{\textit{Department of Computer} \\
% \textit{Science and Engineering} \\
% \textit{Daffodil International University}\\
% Dhaka, Bangladesh \\
% if.firozy880@gmail.com}
% }
\author{
Md Zehan Alam\textsuperscript{1}, Tonmoy Roy\textsuperscript{2}, H.M. Nahid Kawsar\textsuperscript{3}, and Iffat Rimi\textsuperscript{4} \\
\textsuperscript{1}\textit{Department of Mathematics, University of Dhaka, Dhaka, Bangladesh} \\
\textsuperscript{2}\textit{Department of Data Analytics and Information Systems, Utah State University, Logan, Utah, USA} \\
\textsuperscript{3}\textit{Department of Applied Mathematics, Noakhali Science and Technology University, Noakhali, Bangladesh} \\
\textsuperscript{4}\textit{Department of Computer Science and Engineering, Daffodil International University, Dhaka, Bangladesh} \\
Email: zehan.alamm@gmail.com, tonmoy.roy@usu.edu, nahidkawsarnhd3@gmail.com, if.firozy880@gmail.com
}

\maketitle
\thispagestyle{firstpage}

\begin{abstract}
This paper explores and enhances the application of Transfer Learning (TL) for multilabel image classification in medical imaging, focusing on brain tumor class and diabetic retinopathy stage detection. The effectiveness of TL—using pre-trained models on the ImageNet dataset—varies due to domain-specific challenges. We evaluate five pre-trained models—MobileNet, Xception, InceptionV3, ResNet50, and DenseNet201—on two datasets: Brain Tumor MRI and APTOS 2019. Our results show that TL models excel in brain tumor classification, achieving near-optimal metrics. However, performance in diabetic retinopathy detection is hindered by class imbalance. To mitigate this, we integrate the Synthetic Minority Over-sampling Technique (SMOTE) with TL and traditional machine learning(ML) methods, which improves accuracy by 1.97\%, recall (sensitivity) by 5.43\%, and specificity by 0.72\%. These findings underscore the need for combining TL with resampling techniques and ML methods to address data imbalance and enhance classification performance, offering a pathway to more accurate and reliable medical image analysis and improved patient outcomes with minimal extra computation powers.
\end{abstract}


\section{Introduction}
Medical image classification is crucial in the field of healthcare, where accurate diagnosis plays a pivotal role in patient outcomes. The advent of Convolutional Neural Networks (CNNs) has significantly advanced this field, enabling automated systems to classify diseases like Lung Cancer \cite{b6}, Breast Cancer \cite{b5} \cite{b7}, Diabetic Foot Ulcer \cite{b7} and many more with remarkable accuracy. However, developing highly accurate models is challenging, particularly when dealing with complex datasets that may be small \cite{b9} or imbalanced.

Transfer Learning (TL) offers a solution by leveraging pre-trained models on extensive, varied datasets, improving their adaptability to specialized tasks like medical image classification \cite{b8}. This study focuses on applying TL to two distinct medical datasets: brain tumor MRI scans \cite{b1} and diabetic retinopathy retinal images \cite{b2}, both of which present distinct challenges. The brain tumor dataset is relatively balanced, whereas the diabetic retinopathy dataset is highly imbalanced, mirroring real-world clinical scenarios.

Using CNN architectures like MobileNet, Xception, InceptionV3, ResNet50, and DenseNet201, this research investigates the effectiveness of TL in addressing these challenges. Additionally, traditional ML methods, combined with SMOTE, are integrated to tackle class imbalance, aiming to improve model accuracy, particularly in underrepresented classes.

This study contributes to the expanding field of TL in medical imaging, providing empirical insights into the performance of various pre-trained models and demonstrating the potential benefits of combining TL with ML techniques and SMOTE. This approach enhances diagnostic accuracy while remaining computationally efficient, paving the way for broader applications of TL in medical imaging and fostering advancements in automated medical diagnosis.

\section{Literature Review}

\subsection{CNN for Medical Image Classification}
CNNs have proven highly effective in medical image classification due to their capability to learn hierarchical features directly from raw image data, leading to enhanced accuracy in detecting complex patterns. For example, Saleh \textit{et al.} combined CNNs with Support Vector Machines (SVM) for lung cancer classification in CT images, achieving better results than traditional methods \cite{b6}. Yang \textit{et al.} used a two-stage selective ensemble of CNNs to improve classification accuracy on both CIFAR-10 and medical datasets \cite{b4}. Ul Haq \textit{et al.} proposed a deep CNN approach incorporating feature fusion and ensemble learning to enhance mammographic abnormality detection \cite{b5}. Rahman \textit{et al.}  developed a model integrating U-Net and YOLO for the automatic detection and localization of breast lesions in mammography \cite{b3}. These studies collectively highlight CNNs' adaptability and effectiveness in various medical image classification tasks.

\subsection{Transfer Learning in Medical Image Classification}
Transfer Learning, which applies pre-trained CNNs to medical imaging, is a powerful tool for improving classification accuracy. G. Ayana \textit{et al.} reviewed its use in breast cancer diagnosis, noting the effectiveness of models like AlexNet, VGGNet, GoogLeNet, ResNet, and Inception \cite{b7}. L. Alzubaidi \textit{et al.} compared Transfer Learning's success in diabetic foot ulcer classification to animal classification tasks, demonstrating its versatility \cite{b8}. D. Müller \textit{et al.} assessed ensembles of pre-trained models across different datasets, highlighting Transfer Learning's capacity to enhance performance even with limited medical data \cite{b9}. Transfer Learning has shown strong potential in medical imaging by enhancing classification accuracy with limited data through pre-trained models.

\subsection{Transfer Learning in Brain Tumor Detection}
Transfer Learning has proven effective in brain tumor detection. R. Haque \textit{et al.} introduced NeuroNet19, a deep neural network using VGG19 and a novel Inverted Pyramid Pooling Module (iPPM) for MRI-based tumor detection  gaining 99.3\% accuracy \cite{b10}. J. Kang \textit{et al.} utilized pre-trained CNNs to extract features from MRI scans, which were then classified using ML algorithms to predict tumor presence with 93.72\% accuracy \cite{b11}. N. Ullah \textit{et al.} assessed nine pre-trained classifiers, including InceptionV3 and Xception, for automatic brain tumor detection, showcasing the adaptability of Transfer Learning in improving diagnostic accuracy upto 98.91\% \cite{b12}.

\subsection{Transfer Learning in Diabetic Retinopathy Stage Detection}
Transfer Learning has been effectively used for Diabetic Retinopathy (DR) stage detection by leveraging pre-trained models from ImageNet. D. C. R. Novitasari \textit{et al.} achieved 98.2\% accuracy on the MESSIDOR dataset using CNNs combined with Deep Extreme Learning Machine (DELM), while M. K. Jabbar \textit{et al.} attained 96.61\% accuracy on the EYEPACS dataset by leveraging pre-trained VGGNet for feature extraction and transfer learning \cite{b13, b14}. Unlike these datasets APTOS 2019 dataset is much smaller and in varied lighting condition. C. Lahmar \textit{et al.} achieved accuracy of 93.09\% by integrating SVM with MobileNetV2 but for 2 class DR classification, making the dataset balanced \cite{b15}. On the other hand, N. Sikder \textit{et al.} achieved 91\% accuracy on the complex and imbalanced APTOS dataset, comprising five classes, by employing meticulous image preprocessing and ensemble learning techniques \cite{b1}.

\section{Research Methodology}

\begin{figure}[htbp]
\centerline{\includegraphics[width=25em]{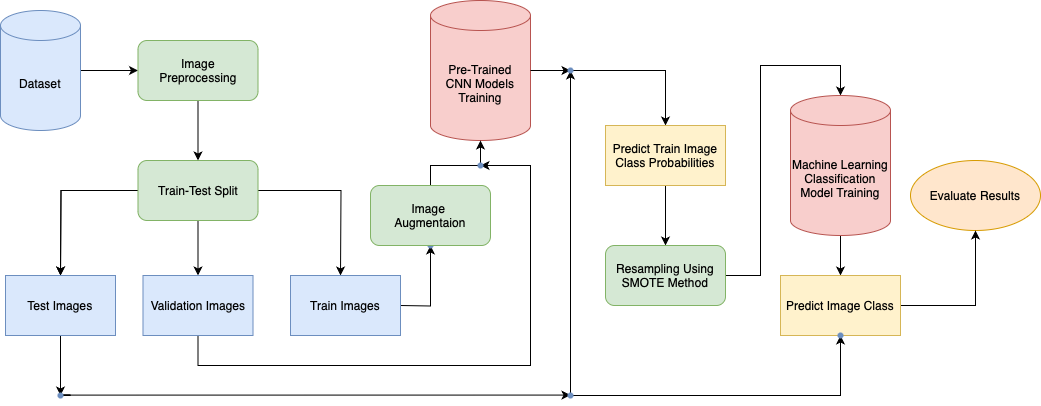}}
\caption{Workflow of the Proposed Model for Medical Image Classification}
\label{fig:lifecycle}
\end{figure}

We employed transfer learning techniques on two diverse medical datasets for image classification. As illustrated in Figure \ref{fig:lifecycle}, our approach involved using pre-trained CNN models to generate class predictions, followed by resampling to address dataset imbalance. These balanced datasets were then utilized to train machine learning models, which were evaluated empirically to determine their predictive performance.

\subsection{Datasets}\label{AA}
\begin{figure}[htbp]
\centerline{\includegraphics[width=22em]{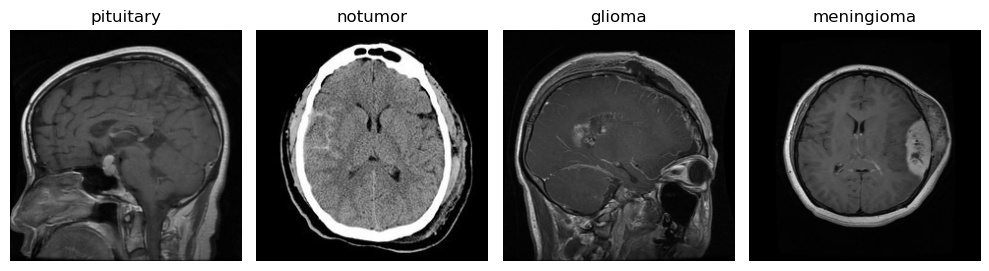}}
\caption{Sample Images from Each Class of BT Dataset}
\label{fig:BT-images}
\end{figure}

\begin{figure}[htbp]
\centerline{\includegraphics[width=18em]{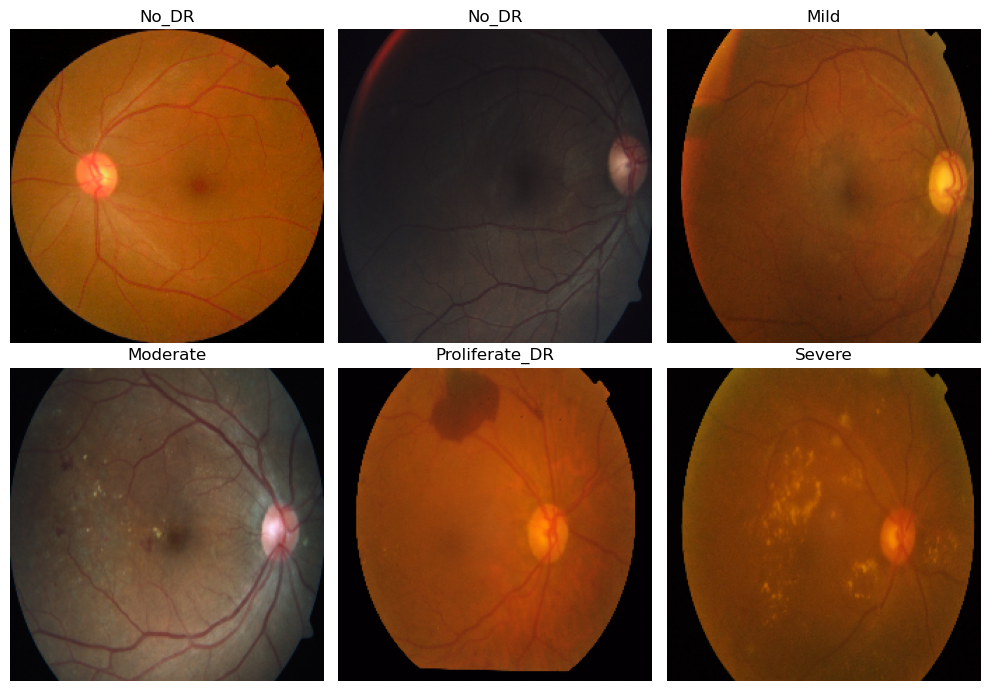}}
\caption{Sample Images from Each Class of DR Dataset}
\label{fig:DR-images}
\end{figure}

For our study, we utilized two publicly available datasets from Kaggle: the 'Brain Tumor MRI' (BT) dataset \cite{b1} and the 'APTOS 2019 Blindness Detection' (DR) dataset \cite{b2}. The BT dataset, sourced from Figshare, SARTAJ, and Br35H, contains 7,023 MRI images of human brains (Figure \ref{fig:BT-images}). In contrast, the DR dataset includes 3,662 retinal images (Figure \ref{fig:DR-images}) obtained via fundus photography under various imaging conditions. Each image has been rated by a clinician for the severity of diabetic retinopathy.

\begin{table}[htbp]
\caption{Distribution of Classes for BT and DR Datasets.}
\begin{center}
\begin{tabular}{|l|c|c|}
\hline
\textbf{Dataset} & \textbf{Class Name} & \textbf{Number of Images} \\
\hline
\multirow{4}{*}{BT} 
    & Glioma & 1,621 \\
    & Meningioma & 1,645 \\
    & No Tumor & 2,000 \\
    & Pituitary Tumor & 1,757 \\
\hline
\multirow{5}{*}{DR} 
    & No\_DR & 1,805 \\
    & Mild & 370 \\
    & Moderate & 999 \\
    & Severe & 193 \\
    & Proliferative\_DR & 295 \\
\hline
\end{tabular}
\end{center}
\label{tab:Dataset-Details}
\end{table}

Notably, the APTOS dataset is highly imbalanced, with the 'No\_DR' class constituting 49.5\% of the total images. This imbalance reflects real-world conditions, where cases of no diabetic retinopathy are more common, and the dataset is characterized by variability in imaging conditions and class distribution.

For the training of CNN models, it is essential to split the data into training, validation, and test sets. The BT Dataset already provided a predefined split, with almost 80\% of the images allocated for training. I further divided the remaining 20\% of the dataset into equal parts, using half for validation and the other half for testing. For the DR dataset, I partitioned 75\% of the images for training, with the remaining 25\% evenly split between validation and test sets.

\subsection{Data Preprocessing and Augmentation}
Both datasets contain images with varying aspect ratios and sizes. To ensure consistency in our processing pipeline, we resized all images from both datasets to 299 × 299 pixels and applied normalization to scale pixel values from the range [0, 255] to [0, 1]. Following preprocessing, we utilized the ImageDataGenerator class from the scikit-learn library to perform data augmentation. The ImageDataGenerator class is particularly advantageous as it enables real-time data augmentation, generating augmented images on-the-fly during model training. The specific augmentations applied included brightness adjustment (brightness\_range=(0.8, 1.2)), horizontal and vertical flipping, nearest-neighbor filling, and a rotation range of up to 60 degrees. These augmentation techniques were employed to enhance model generalization and reduce the risk of overfitting by providing a diverse set of training examples.

\subsection{Convolutional Neural Networks (CNNs)}
CNNs are deep neural networks tailored for visual data analysis, consisting of several layers that each play a role in feature extraction and processing from images. Key layers in a typical CNN include:

\begin{itemize}
    \item \textbf{Convolutional Layers:} These apply filters to the input image, capturing spatial patterns and hierarchies.
    \item \textbf{Activation Layers:} Activation functions such as ReLU (Rectified Linear Unit) add non-linearity to the model.
    \item \textbf{Pooling Layers:} Operations like max pooling reduce the feature map dimensions while preserving crucial information and decreasing computational load.
    \item \textbf{Fully Connected Layers:} These layers perform high-level reasoning and classification based on the features extracted by the previous layers.
\end{itemize}

\subsection{Transfer Learning}

Transfer learning is a machine learning technique where a model trained on one task is adapted to perform a different but related task. This approach is particularly beneficial when the target task has limited data available. The core idea is to leverage knowledge gained from a pre-trained model, typically trained on a large and diverse dataset such as ImageNet, to enhance performance on the target task.

We employed pre-trained models, excluding their top classification layers, and adapted them with custom layers to suit our specific task. We flattened the output from each pre-trained model and appended a dense layer with 128 units and `ReLU' activation. To reduce overfitting, Dropout layers were strategically added with rates of 30\% and 25\%.  The final layer used `softmax' activation to generate class probabilities.

The models were compiled using the `Adamax` optimizer with a learning rate of 0.001 and `categorical crossentropy` loss, evaluated via accuracy, precision, and recall. Training incorporated the `ReduceLROnPlateau` callback with parameters `monitor='val\_loss'`, `factor=0.3`, `patience=2`, `verbose=2`, and `mode='min'`, which reduced the learning rate after two epochs without improvement in validation loss. Additionally, the `EarlyStopping` callback with `monitor='val\_loss'`, `min\_delta=0.01`, and `patience=3` halted training if validation loss failed to improve. These strategies ensured stable convergence over 25 epochs. Predicted class probabilities for training and test datasets were retained for subsequent pipeline use.

Several pre-trained CNN architectures from the ImageNet dataset are commonly used for transfer learning due to their proven performance. The models utilized in this study are:

\subsubsection{MobileNet}

MobileNet is a compact CNN designed for mobile and embedded systems. It uses depthwise separable convolutions, dividing the convolution process into depthwise and pointwise operations, thus reducing computational demands. This makes it suitable for medical image classification under resource constraints.

\subsubsection{Xception}

Xception extends depthwise separable convolutions across its layers, incorporating residual connections to improve feature extraction and gradient flow. Its capability to capture detailed patterns enhances its effectiveness for complex medical images.

\subsubsection{InceptionV3}

InceptionV3 employs Inception modules to analyze images at multiple scales, combining various convolutional filters and pooling techniques. This multi-scale approach is well-suited for handling the diverse and intricate features found in medical images.

\subsubsection{ResNet50}

ResNet50, with 50 layers, uses residual blocks with skip connections to address the vanishing gradient issue, enabling deeper network training. This depth allows the model to capture complex features necessary for accurate medical image classification.

\subsubsection{DenseNet201}

DenseNet201 extends the DenseNet architecture to 201 layers, enhancing feature reuse and gradient flow. This deeper structure excels in learning intricate patterns in medical images, making it highly effective for detailed feature extraction tasks.

\subsection{Resampling Method}
To enhance model performance, we used class probabilities from the CNN's training images as features, with the original labels as target variables. This process enabled us to form a new dataset for training traditional machine learning algorithms. The resulting dataset retains the distribution of the original data, which is challenging with the DR dataset due to significant class imbalance.

To tackle this imbalance, we applied SMOTE, a common resampling technique that generates synthetic samples for the minority class by interpolating between existing instances. SMOTE helps create a more balanced dataset, improving the performance and robustness of machine learning models when dealing with imbalanced data.

\subsection{Machine Learning Algorithms}
In the final phase of our study, we applied a variety of traditional machine learning classification models to predict outcomes on the resampled datasets, which used 'balanced' class\_weight parameter if available. To ensure a robust and comprehensive performance evaluation, we utilized a diverse set of algorithms. This included ensemble methods such as Voting Classifier(which combined Logistic Regression, Linear SVM, SGDClassifier, Extra Trees Classifier, and Random Forest Classifier). Other ensemble techniques employed were Random Forest, Extra Trees, Hist Gradient Boosting, XGBoost, Bagging, Gradient Boosting, CatBoost, and LightGBM. Additionally, we evaluated linear models including Logistic Regression, Linear SVM, and SGD, as well as the Naive Bayes model for probabilistic classification. These models were selected to thoroughly assess their performance on the balanced datasets.

\subsection{Evaluation Metrics}

We employed several key metrics to evaluate the performance of our models in multiclass classification tasks. These metrics are crucial for assessing the effectiveness of models in the medical domain, where accurate predictions are vital for diagnosis and treatment.

\subsubsection*{Confusion Matrix}
The confusion matrix provides a detailed breakdown of model performance across all classes, showing true positives (TP), true negatives (TN), false positives (FP), and false negatives (FN). It helps in understanding the distribution of predictions across different classes and is essential for evaluating model performance.

\begin{table}[htbp]
\caption{Confusion Matrix for Multiclass Classification}
\begin{center}
\begin{tabular}{|c|c|c|c|}
\hline
 \diagbox{Actual}{Predicted} & \textbf{Class 1} & \textbf{Class 2} & \textbf{... Class N} \\
\hline
\textbf{Class 1} & $TP_1$ & $FN_{12}$ & ... \\
\hline
\textbf{Class 2} & $FP_{21}$ & $TP_2$ & ... \\
\hline
\textbf{... Class N} & ... & ... & $TP_N$ \\
\hline
\end{tabular}
\end{center}
\label{tab:confusion-matrix-multiclass}
\end{table}

\subsubsection*{Accuracy}
Accuracy measures the proportion of correct predictions across all classes. It is calculated as:

\begin{equation}
\text{Accuracy} = \frac{\sum_{i=1}^{N} TP_i}{\sum_{i=1}^{N} (TP_i + TN_i + FP_i + FN_i)}
\label{eq:accuracy-multiclass}
\end{equation}

Accuracy provides an overall performance measure of the model but may be less informative in the presence of class imbalance.

\subsubsection*{Precision}
Precision reflects the proportion of correctly identified positive instances among all positive predictions, averaged using the 'micro' method:

\begin{equation}
\text{Precision} = \frac{\sum_{i=1}^{N} TP_i}{\sum_{i=1}^{N} (TP_i + FP_i)}
\label{eq:precision-multiclass}
\end{equation}

Precision is critical for minimizing false positives, which helps in avoiding unnecessary treatments or interventions.

\subsubsection*{Recall (Sensitivity)}
Recall measures the model's ability to identify all positive instances, averaged using the 'micro' method:

\begin{equation}
\text{Recall} = \frac{\sum_{i=1}^{N} TP_i}{\sum_{i=1}^{N} (TP_i + FN_i)}
\label{eq:recall-multiclass}
\end{equation}

Recall is essential for ensuring that conditions are not missed, thereby reducing false negatives.

\subsubsection*{F1-Score}
The F1-Score balances precision and recall by computing their harmonic mean:

\begin{equation}
\text{F1-Score} = 2 \cdot \frac{\text{Precision} \cdot \text{Recall}}{\text{Precision} + \text{Recall}}
\label{eq:f1-score-multiclass}
\end{equation}

The F1-Score provides a balance between precision and recall, which is crucial for managing both false positives and false negatives.

\subsubsection*{Specificity (True Negative Rate)}
Specificity measures the proportion of correctly identified negatives:

\begin{equation}
\text{Specificity} = \frac{\sum_{i=1}^{N} TN_i}{\sum_{i=1}^{N}(TN_i + FP_i)}
\label{eq:specificity-multiclass}
\end{equation}

Specificity is important for ensuring that healthy individuals are correctly classified, avoiding false positives.

\section{Results and Analysis}

The models were evaluated in two phases. Initially, we used pre-trained models with transfer learning to predict the training data and select the optimal model for feature creation. The chosen model was then applied to the respective dataset, and various machine learning models were compared on the features generated by these pre-trained models. We'll only use the top 5 of the final results to evaluate. The model training and testing were conducted on the Kaggle platform, utilizing an NVIDIA Tesla P100 GPU, 4 CPU cores, and 30 GB of RAM.

\subsection{BT Dataset}

\begin{table}[htbp]
\caption{Pre-trained Model Performance for BT dataset: Accuracy (Ac), Precision (Pr), Recall (Rc), F1 Score (F1), Specificity (Sp).}
\begin{center}
\begin{tabular}{|l|c|c|c|c|c|}
\hline
\textbf{Model Name} & \textbf{Ac (\%)} & \textbf{Pr (\%)} & \textbf{Rc (\%)} & \textbf{F1 (\%)} & \textbf{Sp (\%)} \\
\hline
MobileNet & 99.85 & 99.88 & 99.84 & 99.86 & 99.94 \\
\hline
Xception & 99.39 & 99.35 & 99.34 & 99.34 & 99.80 \\
\hline
InceptionV3 & 99.39 & 99.35 & 99.33 & 99.34 & 99.80 \\
\hline
ResNet50 & 99.24 & 99.18 & 99.17 & 99.18 & 99.75 \\
\hline
DenseNet201 & 87.35 & 87.32 & 86.49 & 86.46 & 95.78 \\
\hline
\end{tabular}
\end{center}
\label{tab:BT-CNN-performance}
\end{table}

For the BT dataset, as shown in Table~\ref{tab:BT-CNN-performance}, the MobileNet model demonstrated superior performance across all metrics with an accuracy (Ac) of 99.85\%, precision (Pr) of 99.88\%, recall (Rc) of 99.84\%, F1 Score (F1) of 99.86\%, and specificity (Sp) of 99.94\%. Other models exhibited similar results with only a 0.4\% difference in accuracy. Transfer learning proved highly effective for this dataset, achieving near-perfect scores. 

\begin{table}[htbp]
\caption{ML Model Performance for BT dataset: Accuracy (Ac), Precision (Pr), Recall (Rc), F1 Score (F1), Specificity (Sp).}
\begin{center}
\begin{tabular}{|l|c|c|c|c|c|}
\hline
\textbf{Model Name} & \textbf{Ac (\%)} & \textbf{Pr (\%)} & \textbf{Rc (\%)} & \textbf{F1 (\%)} & \textbf{Sp (\%)} \\
\hline
Logistic Reg. & 99.85 & 99.88 & 99.84 & 99.85 & 99.94 \\
\hline
XGBoost & 99.85 & 99.88 & 99.84 & 99.85 & 99.94 \\
\hline
Random Forest & 99.85 & 99.88 & 99.84 & 99.85 & 99.94 \\
\hline
Gradient Boosting & 99.85 & 99.88 & 99.84 & 99.85 & 99.94 \\
\hline
Extra Trees & 99.85 & 99.88 & 99.84 & 99.85 & 99.94 \\
\hline
\end{tabular}
\end{center}
\label{tab:BT-ML-performance}
\end{table}

After applying SMOTE to the predicted data and training the ML models, the performance on the test data, as shown in Table~\ref{tab:BT-ML-performance}, mirrored MobileNet's performance. This indicates that SMOTE did not enhance performance at all. 

\begin{figure}[htbp]
\centerline{\includegraphics[width=18em,height=14em]{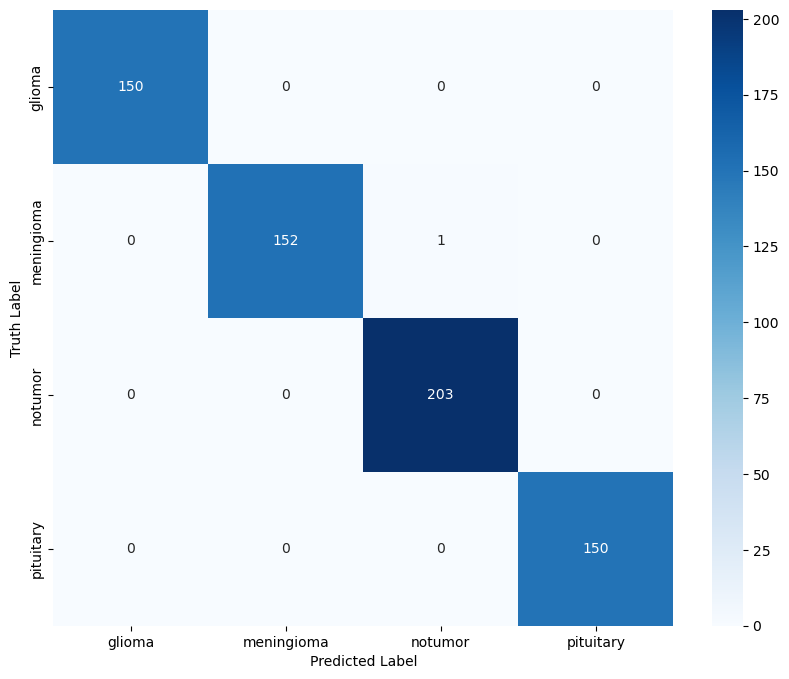}}
\caption{Confusion Matrix of Final Prediction on BT Dataset}
\label{fig:CM-BT}
\end{figure}

The confusion matrix in Figure~\ref{fig:CM-BT} shows a single misclassification: a meningioma tumor was incorrectly labeled as no tumor, which could have significant consequences. However, the model correctly classified all other test images.

\subsection{DR Dataset}

\begin{table}[htbp]
\caption{Pre-trained Model Performance for DR dataset: Accuracy (Ac), Precision (Pr), Recall (Rc), F1 Score (F1), Specificity (Sp).}
\begin{center}
\begin{tabular}{|l|c|c|c|c|c|}
\hline
\textbf{Model Name} & \textbf{Ac (\%)} & \textbf{Pr (\%)} & \textbf{Rc (\%)} & \textbf{F1 (\%)} & \textbf{Sp (\%)} \\
\hline
Xception & 90.17 & 83.19 & 80.06 & 81.00 & 97.46 \\
\hline
DenseNet201 & 89.30 & 80.24 & 75.59 & 77.48 & 97.32 \\
\hline
InceptionV3 & 87.55 & 76.64 & 74.35 & 74.73 & 96.98 \\
\hline
MobileNet & 78.82 & 72.90 & 49.46 & 48.02 & 94.20 \\
\hline
ResNet50 & 27.73 & 25.66 & 20.09 & 9.00 & 80.21 \\
\hline
\end{tabular}
\end{center}
\label{tab:DR-CNN-performance}
\end{table}

For the DR dataset, Table~\ref{tab:DR-CNN-performance} highlights that Xception outperformed other pre-trained models with an accuracy of 90.17\%, precision of 83.19\%, recall/sensitivity of 80.06\%, F1 Score of 81.00\%, and specificity of 97.46\%. Unlike the BT dataset, performance variations among models were more pronounced. After Xception, DenseNet201, and InceptionV3 achieved high accuracy, but all models exhibited poor recall (sensitivity). MobileNet and ResNet50 performed poorly overall. As a result, Xception was chosen for subsequent steps.

\begin{table}[htbp]
\caption{ML Model Performance for DR dataset: Accuracy (Ac), Precision (Pr), Recall (Rc), F1 Score (F1), Specificity (Sp).}
\begin{center}
\begin{tabular}{|l|c|c|c|c|c|}
\hline
\textbf{Model Name} & \textbf{Ac (\%)} & \textbf{Pr (\%)} & \textbf{Rc (\%)} & \textbf{F1 (\%)} & \textbf{Sp (\%)} \\
\hline
Voting Classifier & 92.14 & 82.10 & 85.50 & 83.52 & 98.18 \\
\hline
SGD & 91.70 & 81.39 & 84.80 & 82.68 & 98.07 \\
\hline
Linear SVM & 91.48 & 81.00 & 84.46 & 82.32 & 98.02 \\
\hline
Logistic Reg. & 91.48 & 81.13 & 84.77 & 82.65 & 98.02 \\
\hline
Naive Bayes & 90.17 & 80.43 & 83.20 & 80.44 & 97.70 \\
\hline
\end{tabular}
\end{center}
\label{tab:DR-ML-performance}
\end{table}

After applying SMOTE and evaluating different ML classifiers, as shown in Table~\ref{tab:DR-ML-performance}, the Voting Classifier achieved the best results with an accuracy of 92.14\%, precision of 82.10\%, recall of 85.50\%, F1 Score of 83.52\%, and specificity of 98.18\%. This represents a 1.97\% improvement in accuracy and a 5.44\% increase in sensitivity, with a slight increase in specificity. These improvements are significant, especially in medical imaging where sensitivity and specificity are crucial. 

\begin{figure}[htbp]
\centerline{\includegraphics[width=18em,height=14em]{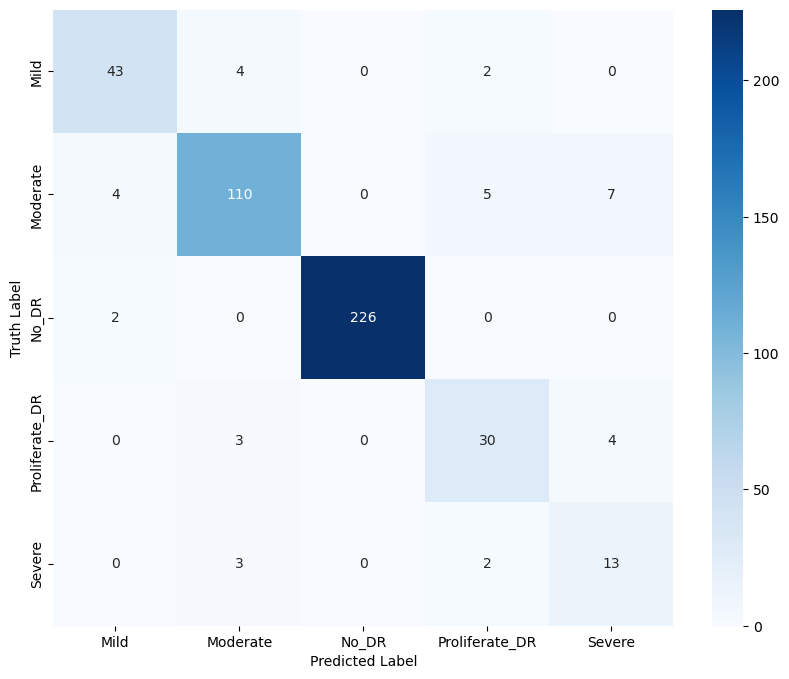}}
\caption{Confusion Matrix for Final Prediction in DR Dataset}
\label{fig:cm-DR}
\end{figure}

The confusion matrix in Figure~\ref{fig:cm-DR} indicates that no images were falsely labeled as no diabetic retinopathy when they actually had diabetes, demonstrating the effectiveness of the chosen model and methods. The moderate class was predominantly misclassified, particularly as Proliferate and Severe classes. However, this misclassification is less critical, as patients with moderate stage DR still require medical intervention.

\section{Discussion}

This study examines the effectiveness of TL in medical image classification, focusing on BT and DR datasets. While promising, TL faces distinct challenges in medical imaging due to the specialized nature of the images and the prevalence of limited, imbalanced datasets.

For the balanced BT dataset, TL models such as MobileNet, Xception, InceptionV3, ResNet50, and DenseNet201 performed exceptionally well, demonstrating the effectiveness of TL when ample annotated data is available. High specificity in medical image classification helps avoid unnecessary treatments, while high sensitivity ensures accurate tumor detection and informs clinical decisions. These metrics were nearly optimal for the BT dataset models. However, the DR dataset, characterized by significant class imbalance, revealed TL's limitations. Despite Xception’s strong performance, the overall accuracy and sensitivity remained inadequate, emphasizing the need for additional strategies to address imbalanced datasets and the specific demands of medical imaging.

To address these issues, we applied the SMOTE method alongside traditional machine learning classifiers, leveraging TL model predictions as features. This hybrid approach significantly improved sensitivity and accuracy for the DR dataset, with the Voting Classifier showing notable gains. However, for the balanced BT dataset, SMOTE provided minimal benefit, indicating that TL alone is sufficient for well-balanced datasets. These findings underscore the importance of tailoring methodologies to dataset characteristics when applying TL to medical imaging. While TL is highly effective, complementary methods such as resampling and ensemble techniques may be required for imbalanced datasets.

\section{Conclusion}

This study underscores the effectiveness of advanced machine learning techniques for medical image classification, offering enhancements to transfer learning methods to address dataset imbalances while maintaining low computational demands. Results demonstrated that TL models, particularly MobileNet, excelled on the balanced BT dataset, achieving high accuracy, precision, and recall. On the imbalanced DR dataset, the integration of the Voting Classifier with SMOTE significantly improved sensitivity and accuracy. These findings highlight the value of combining advanced algorithms with resampling techniques to manage imbalanced datasets and enhance predictive performance. Importantly, this approach is versatile and applicable across various medical imaging tasks, addressing the common challenge of underrepresented classes in medical datasets.

The study emphasizes the need for a comprehensive approach that combines TL with additional methods to handle diverse medical imaging challenges effectively. A notable limitation of this research is the lack of extensive image preprocessing, which could have enhanced feature extraction and potentially improved all metrics. This study primarily concentrated on mitigating class imbalance to enhance performance. While modern techniques like GANs could be beneficial, they are computationally intensive. Future research should investigate more thorough preprocessing techniques in conjunction with this method and assess their application across a broader range of medical imaging contexts. By adopting these strategies, researchers and clinicians can achieve more accurate and reliable predictions, ultimately enhancing diagnostic precision and patient outcomes.

\section*{Acknowledgment}
We acknowledge the valuable contributions of generative AI tools, such as ChatGPT \cite{b16}, for their assistance in proofreading, formatting and coding support.

\end{document}